# Detection of ethanol in water by electrical impedance spectroscopy and resonant perturbation method

Angelo Leo[1, 2*,] **Anna Grazia Monteduro**[1, 2*], Silvia Rizzato[1, 2], Angelo Milone[1, 2] and Giuseppe Maruccio[1, 2]

[1] Department of Mathematics and Physics "Ennio De Giorgi", University of Salento, Omnics Research Group, Via per Monteroni, 73100 Lecce, Italy;
[2] Institute of Nanotechnology, CNR-Nanotec, Omnics Research Group, Via per Monteroni, 73100 Lecce, Italy;
*A.L. and A.G.M. have equally contributed to this work.

### Abstract

Development of highly sensitive, portable and low-cost sensors for the evaluation of ethanol content in liquid is particularly important in several process monitoring, from food to pharmaceutical industry. In this respect, here we report on the design, fabrication and performances of two simple electrical devices, namely interdigitated (ID) electrodes and complementary double split ring resonator (CDSRR), for the detection of ethanol in water though impedance and perturbation resonance methods, respectively. Both sensors resulted to be efficient for detection of low amount of ethanol in water, in particular EIS gives possibility to perform broadband evaluation of ethanol concentration in solution, and the employment of resonant cavities allows to achieve very low limit of detection of 0.2v/v%.

## 1. Introduction

Portable and low-cost sensors for the detection of chemicals in liquid are finding an ever growing demand in many application fields, such as in food industry [1, 2], medical diagnostics [3-6] and for environmental monitoring of contaminants and pollutants [7].

In this contest, development of sensors for the evaluation of ethanol content in liquid is particularly important for process monitoring in distilleries and breweries, in the manufacture of drugs, plastics, lacquers, polishes, plasticizers and cosmetics [8]. Indeed, it is largely used for its miscibility in water and as solvent in several processes, besides being used as an antiseptic, thanks to its ability to kill microorganisms, denaturing their proteins and dissolving fats. However, it is the automotive sector that requires more: approximately 73% of world's ethanol produced is destined as an additive for fuels, in blends with variable percentages [9]; moreover, the bioethanol (along with biodiesel) is considered one of the most promising alternatives today to fossil fuels and is the main biofuel used as a substitute for vehicles road transport.

Current methods for detection of ethanol in liquids are based on indirect calculation of ethanol content or direct determination, respectively by oscillating density meters [10], and infrared spectroscopy [11, 12], techniques which are time-consuming and costly. On the other side, cheaper, portable devices such as hydrometers have poor resolution and a strong temperature dependency [13], which makes them unsuitable for an accurate monitoring of the ethanol content over the typical time of the fermentation process.

Alternative methods for ethanol detection are based on employment of sensitive hydrogels, whose swelling-shrinking state after contact with ethanol can be interrogated for example by pressure sensor, quartz crystal microbalance or optically, leading to discern ethanol concentration variation of 10% [14]; further state-of-art ethanol sensors are merely optical [15, 16] and electrochemical [17-19]. They show excellent sensitivity, but their use is strongly linked to environmental conditions: optical sensors are too voluminous to be integrated in small devices, while the electrochemical ones are expensive and often suffer high temperatures. RF sensors are more suitable for industrial field with respect to the aforementioned ones since they provide the benefits of low cost, contactless and reusability, while maintaining excellent performances. Among the alternative sensors, electrical impedance spectroscopy (EIS) - based sensors [20] and metamaterial structures such as split-ring resonators and complementary ones prove to be useful for the evaluation of ethanol concentration [21, 22], even by integrating them in a microfluidic platform [23].

Here, we propose and compare the performances of two simple electrical devices, namely interdigitated (ID) electrodes and complementary double split ring resonator (CDSRR), for detecting ethanol presence in water though spectroscopic measurements, in particular through impedance and perturbation resonance methods, respectively.

2. Materials and methods

2.1 Fabrication of Electrochemical Impedance Spectroscopy (EIS) device and experimental setup

The electrochemical impedance spectroscopy (EIS) device employed for ethanol evaluation concentration is composed by a sensing module with interdigitated electrodes and an array of resin 3D printed chambers, which separate each sensing area (Figure 1). More in detail, each sensing module consists of four arrays each one containing four couples of interdigitated microelectrodes as transducers, which have a common reference and four signal pins (Figure 1c). These electrodes, characterized by a finger width and interfinger gap of 7 μm, were fabricated by photolithography and lift off process on glass substrates (EOT) using a Karl Suss MA6 mask aligner and AZ5214B resist and thermal evaporation of Cr/Au (3 nm/35 nm). The array of polymeric chambers (50 μL volume each one) was first designed using AUTOCAD and then realized using the Asiga Max 3D printer. The impedance measurements were carried out at room temperature employing an Agilent E4980A precision LCR meter and a home-made software written in Labview (GM-Multiscan) for the data acquisition. An AC-driving voltage of 100 mV was applied over the frequency range from 100 Hz to 1 MHz. The electrical connections between the EIS chip and the LCR meter were performed fabricating dedicated pads on the EIS chip with a specific pitch to match a commercial connector integrated on a printed circuit board (PCB) (Figure 1a-b).

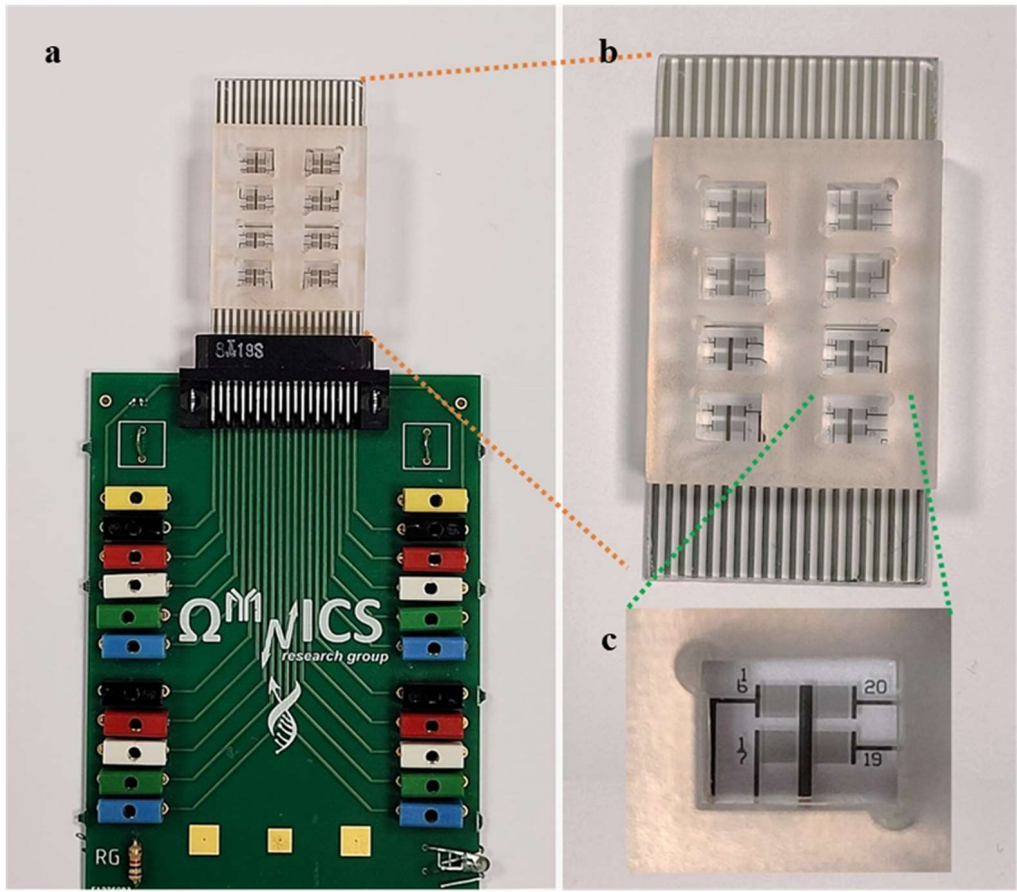

**Figure 1**. (a) EIS sensor with IDT electrodes and 3D resin chamber, connected to the printed circuit board; (b) EIS sensor device, in which the electrodes for external connections are visible on the bottom and on the top of the device and (c) enlargement of a sensor area in which are clearly visible a common reference and four signal pins.

**2.2 Design, simulation and fabrication of CDSRR-based device and experimental setup**

The device exploited for the detection of ethanol concentration in water is a one-port complementary double split ring resonator (CDSRR), realized in a Cu disc of 27 mm diameter and 15 μm height, connected to a right-angle SMA through a metallized plate (the feedline) of dimensions 4 mm × 6 mm, as shown in Figure 2a. A peculiar attention was paid to the lines width design for lossy coupling the devices to instruments and for promote resonance. The lossy coupling regime can be obtained by acting on impedance of entire devices at analytically estimated fundamental modes. In particular a good performance can be ensured by modelling (i) the feed line to have matched impedance ($Z \simeq 50$ Ohm) near connectors and high impedance (~100 Ohm) near resonator, and (ii) the ring to have impedance in the range $50\ \Omega < Z < 100\ \Omega$. By employing KiCad software we extracted the circuit length and line width by setting the impedance conditions and losses in conductor and substrate. The resonator is composed by two concentric rings with mean radii $r_1$ and $r_2$ of 4.7 mm and 5.8 mm respectively, width of 0.8 mm and separated by a coupling gap $c$ of 0.3 mm. Each ring shows a 100 µm slit ($g_1 = g_2$), which lies on the axis joining the centre of the resonator with the signal via. The built-up board is constituted by top and bottom copper layers of 12-18 µm, and a substrate of Improved-FR4 (IS400), which is 1 mm thicker and with permittivity $\varepsilon_r$ ranging from 4.43 to 4.51 for frequencies comprised among 500 MHz to 20 GHz and loss tangent $\delta_t$ of 0.0149÷0.0189 in same spectrum.

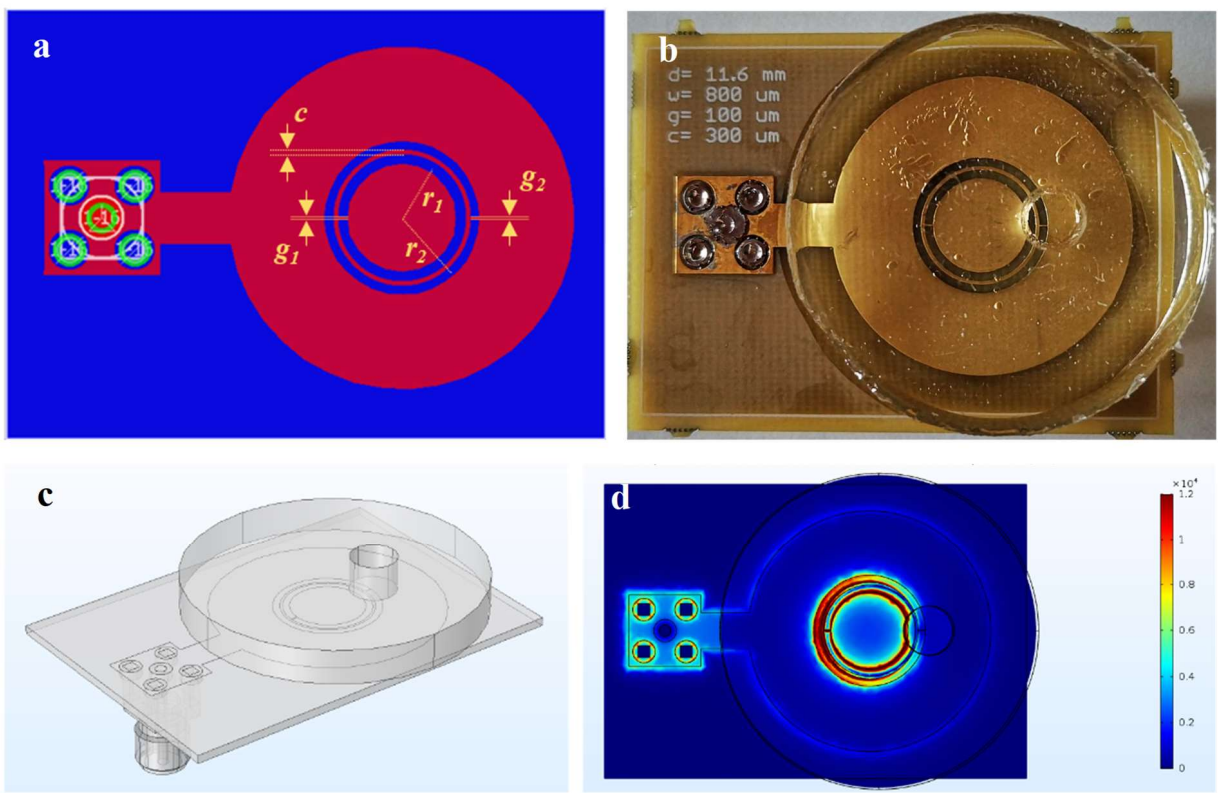

**Figure 2.** (a) CAD project of CDSRR. The upper metallized region is indicated in red, while the ground is represented in blue; vias are shown as green circular crowns. Particular features are indicated in yellow. (b) Device covered by a PDMS disc in which the well was realized. (c) 3D CAD drawn with Comsol© Multiphysics 5.3a for the whole device. (d) Module of the RF electric field at first mode for a slice adjacent to the interface between the resonator and the PDMS disc.

For the ethanol sensing application, the resonator was then covered by a polydimethylsiloxane (PDMS) disk of 4.2 mm height and 35 mm in diameter, in which a hole of 5.5 mm in diameter was practiced to discover the sensing area. The PDMS disc was attached to the device by applying a thin layer of liquid PDMS and placing the system in the oven for 30 minutes at 70°C. In such a way, a well with a total capacity of 100 µl was realized. The whole CDSRR device is shown in Figure 2b.

The sensing experiments were carried out directly connecting the device to a Keysight E5061B ENA Vector Network Analyzer by means of SMA connectors to measure the module of scattering parameter $S_{11}$, thus eliminating distortion effects due to the presence of cables. In these conditions, the resonator shows the first resonance peak at 1.094 GHz, with amplitude of -16.37 dBm, indicating a quality factor $Q \sim 100$ (empty sensor).

To evaluate how the electric field is distributed in proximity of CDSRR and to test the performances of the projected ring before the fabrication, Comsol© Multiphysics 5.3a (RF module) was employed to simulate the reflection spectrum. Therefore, we first designed the CAD of the whole device as reported in Figure 2c, including the PDMS disc with well, the SMA connector and a big sphere to take into account the environment (not shown for clarity). For each material, we inserted the values of electrical conductivity, dielectric constant and magnetic permeability taken from the literature. We found that the first mode, which was experimentally used, is characterized by an eigenfrequency of 1.081 GHz, with a relative discrepancy with respect to measurements less than 2%; the mismatch can be further reduced by using real values of the electrical characteristics. From the simulation on the first mode, it is possible to notice the RF electric field distribution at the interface between the PDMS disc and the ring itself: as reported in Figure 2d, the presence of the well in proximity of the external slit deeply modifies the field distribution, which results to be not symmetric with respect to the centre of ring. Therefore, the presence of a solution inside the well significantly perturbs the resonator, as the dielectric constant of material increases.

## 3. Results and discussion
### 3.1 Detection of ethanol using EIS sensor

In Figure 3 the impedance response of a typical EIS sensor is reported by varying the concentration of ethanol in demineralized (DI) water from 0% to 100%. During the experimental measurements, a resin chamber over a sensing area was filled with 50 µl of sample solution and the impedance data were acquired over the frequency range from 100 Hz to 1 MHz. Then the solution was removed, the well washed with DI water and dried with nitrogen before to record the impedance response of a solution with a different ethanol percentage content. The impedance data are presented as Nyquist plot, in which the imaginary part of impedance ($Z''$) is plotted as a function of the real part ($Z'$). In Figure 3, we observe that increasing the amount of ethanol in the solution the arc diameter increases, reaching the maximum value with absolute ethanol (blue pentagons in the graph) and the minimum with DI water (black squares in the graph). In order to promote a more in deep understanding on how the electric parameters change by varying the ethanol amount in the water solution the complex impedance data ($Z'$, $Z''$) were analysed with the elemental equivalent circuit method. First, we select the most suitable equivalent circuit analysing the complex impedance of water and then we employed the same equivalent circuit to fit the impedance data recorded by varying the ethanol concentration in water. The equivalent circuit which better describe our data is reported as inset in Figure 4a. It consists of a parallel between a constant phase element (CPE) and a resistance ($R$). This is the most common circuit used to describe a not ideal capacitor, characterized by a distribution of relaxation times [24-26]. The impedance of the CPE element, $Z_{CPE}$, is expressed as in following:

$$Z_{CPE} = \frac{1}{Q_0(j\omega)^n}$$

where $Q_0$ is expressed in SHz$^{-n}$, $\omega = 2\pi f$ is the angular frequency, $0 < n < 1$, which corresponds to an ideal capacitor when $n = 1$ and to an ideal resistor when $n = 0$, $j$ is the imaginary unit.

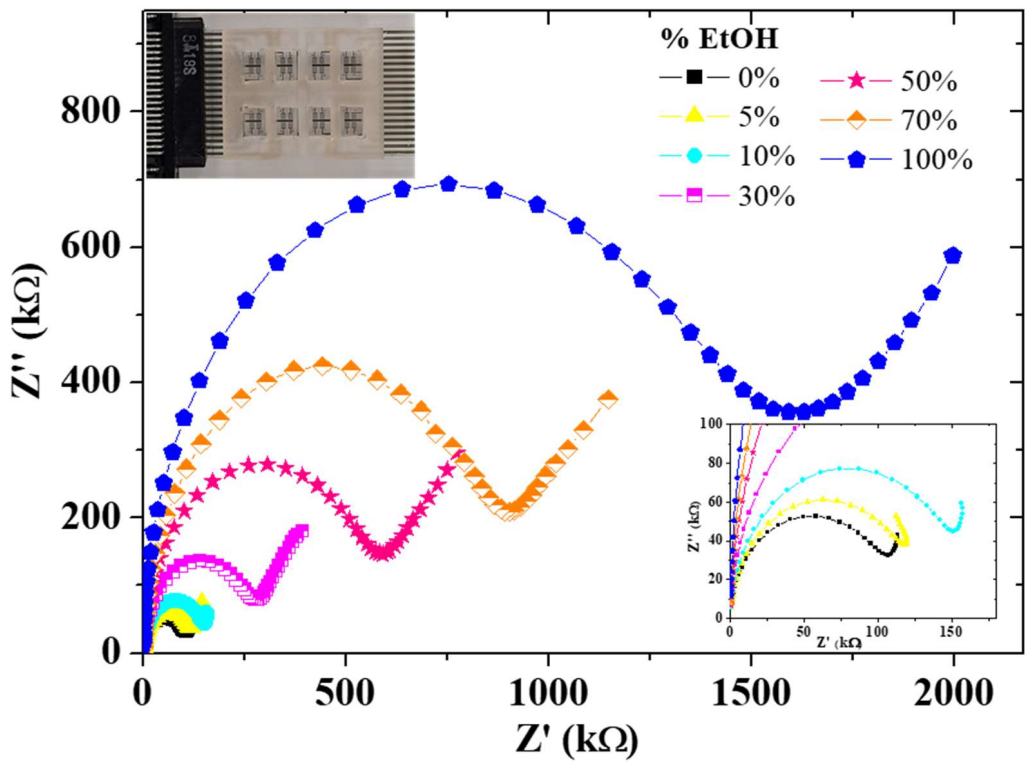

**Figure 3**. Nyquist plots of EIS sensors by varying the %EtOH in DI water. Insets: (top-left) optical photo of the EIS sensor connected to the PCB for impedance measurements; (bottom-down) enlargement of the impedance responses for pure water (black squares) and with low concentration of ethanol, 5% (yellow triangles) and 10% (cyan dots).

To give an idea on how the selected circuit describes our experimental data, we reported in Figure 4a the fit (red line) of the impedance data acquired for a 30% amount of EtOH in solution by using the CPE-R equivalent circuit. The electrical parameter, $Q_0$ and $R$, extracted by fitting the Nyquist plots, were reported in Figure 4b as a function of the %EtOH in the solution. An almost linear trend was observed for $Q_0$ (black dotted line) on the entire EtOH percentage range investigated, while the resistance, $R$ (blue dotted line), shows a variation of the slope for EtOH concentration lower than

30%. Moreover, since $Q_0$ is related to the capacitive properties of the liquid phase, the data obtained suggest a reduction of the dielectric permittivity of the solution by increasing the EtOH content. This is in agreement with the dielectric properties of the pure liquid, since a dielectric constant of about 80 is found in literature for DI water [27, 28] and a permittivity of about 25 for absolute EtOH [29-31]. On the other hand, the increase of the resistance by increasing the ethanol content in the solution agrees with the lower conductivity of ethanol with respect to the water one.

We evaluated the limit of detection (LOD) of the sensor by dividing the $Q_0$ and $R$ uncertainties (calculated considering the average of residuals from the linear fits) with the sensitivity, represented by the slope of the linear fits of the two parameters considered. Specifically, we linear fitted $Q_0$ over the whole percentage variation of ethanol founding a LOD of 2.5% EtOH, while the resistance variation was linear fitted from 0% to 30% EtOH estimating a LOD of 0.7% EtOH.

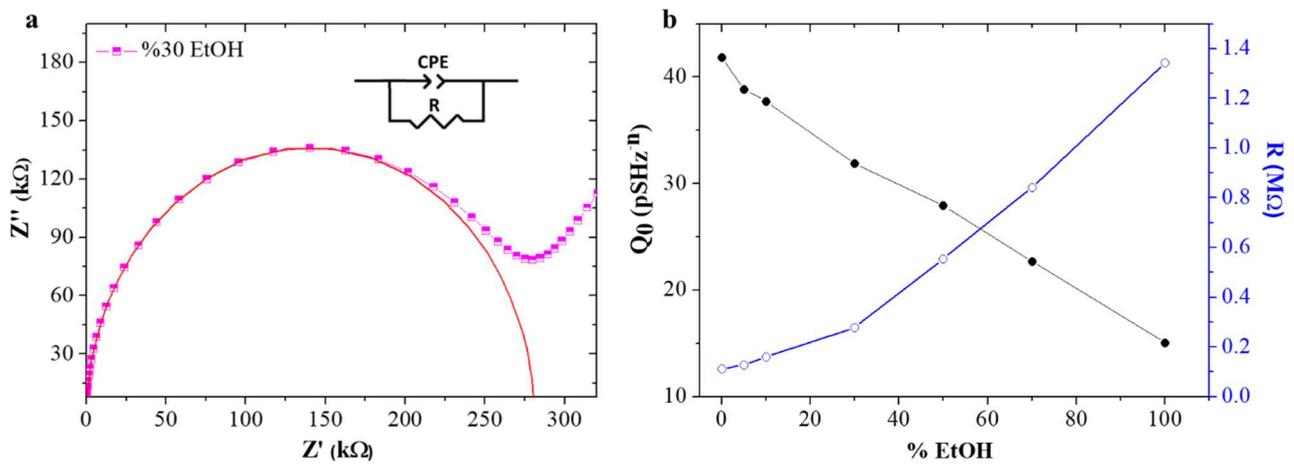

**Figure 4**. (a) Fitting of Nyquist plot recorded for the solution with 30% EtOH (inset: equivalent circuit employed to model the impedance responses acquired for different the ethanol contents). (b) $Q_0$ (black dotted line) and $R$ (blue dotted line) profiles as a function of ethanol concentration.

### 3.2 Detection of ethanol using CDSRR sensor

The CDSRR sensor was tested for the detection of EtOH in liquid phase by depositing in the PDMS well a 100 µl droplet of ethanol-DI water solution. The concentration of ethanol in water was varied from 0% to 100%. As for EIS sensor, after each measurement, the solution was removed, the well was washed with demineralized water and dried with nitrogen flow, bringing the system to its initial conditions, with a maximum frequency shift and a maximum amplitude variation of less than 0.2%. The reflection signals module $S_{11}$ from loaded device are reported in Figure 5a; the loading of device by ethanol/water solution causes a frequency shift and an amplitude variation of the resonant peak, beside a modification of quality factor $Q$ of the resonator. In the Figure 5b the variation of the resonance frequency and $Q$ values as a function of the ethanol concentration are shown. By reducing the concentration of EtOH, a shift of the resonance peak toward lower frequencies was observed, reaching the minimum with water (1.042 GHz), and an increasing of the quality factor $Q$ of the device was also observed, passing from about 100 for absolute EtOH (as for the empty device value) to about 1000 for pure water.

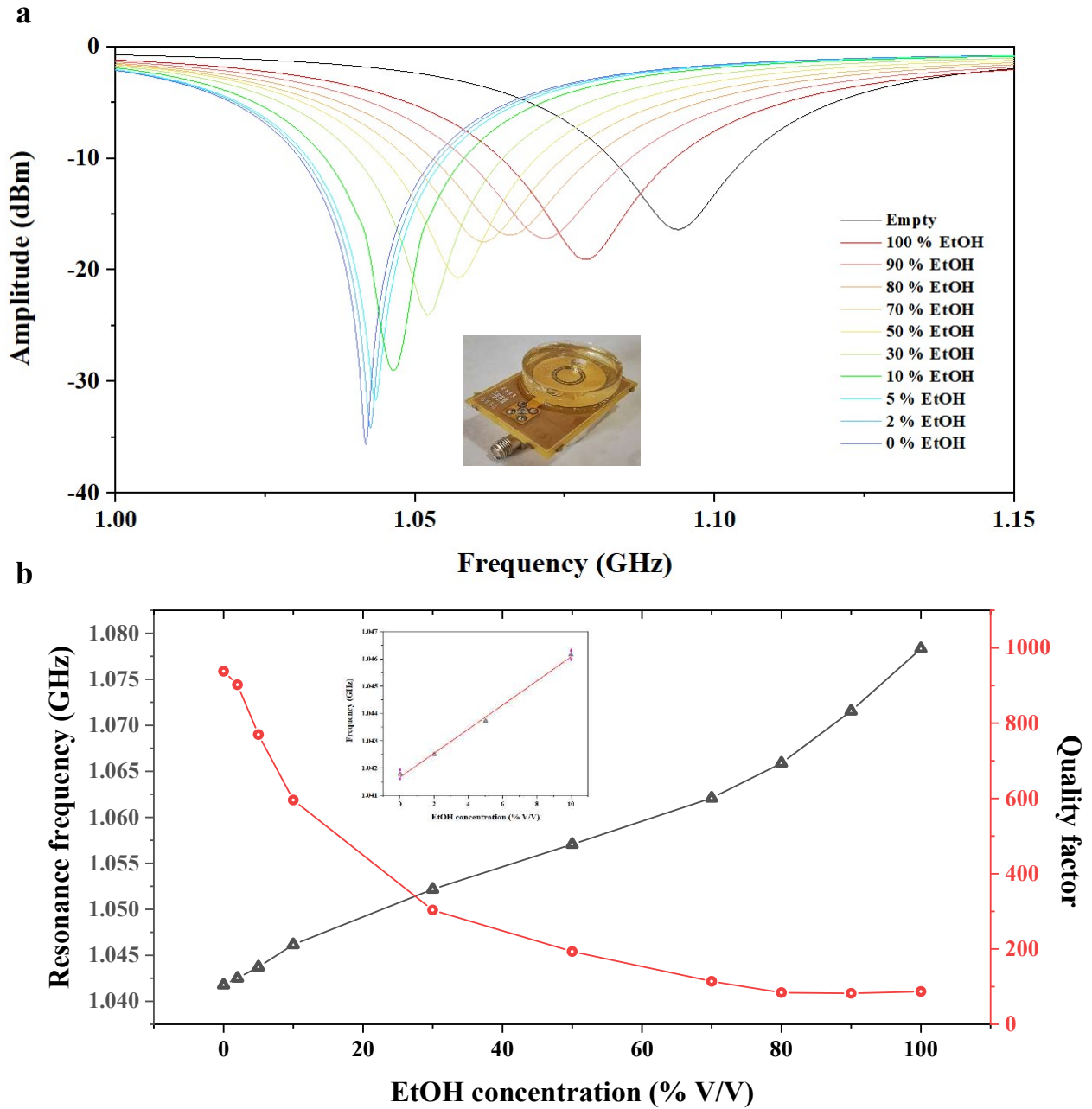

**Figure 5.** (a) Measured $S_{11}$ of water-ethanol mixture for different volume fraction. Inset: optical photo of the CSRR sensor. (b) Resonant frequency and quality factor for different water-ethanol volume fractions. Inset: linearization of the response for the first 4 investigated concentrations.

Moreover, if resonance frequency of the device is almost linear with the increase of EtOH concentration, the quality factor trend shows a minimum when the device is loaded with a solution of 80% EtOH concentration. By fitting the resonance frequencies of the loaded device with respect to the concentration of ethanol in water up to 10% (inset of Figure 5b), we obtained a sensitivity of

0.329 ± 0.016 MHz for 1% increment of EtOH content in water. Estimation of LOD of CSRR is carried out by dividing the resonance frequency uncertainty (calculated, as for EIS sensor, by considering the average of residuals from the linear fit) with the sensitivity, obtaining a LOD of 0.2% of EtOH.

The high sensing performance of our CDSRR proposed sensor is mainly associated to the improved quality factor of the device, which in our case reaches values higher than ring resonators exploited in present literature for the evaluation of ethanol content in water [23, 32-35].

**Conclusions**

In this paper we show the detection of ethanol in water at very low percentage v/v by exploiting two different devices, namely EIS sensor and CDSRR. For each measurement layout, two quantities were investigated: EIS led to evaluate resistance $R$ and the $Q_0$ parameter related respectively to the electrical conductivity and dielectric properties of the investigated solution, while perturbation of ring resonator gave access to working eigenfrequency shift and quality factor variation. Both the devices and related methods resulted efficient to the detection of low amount of ethanol in water, however while EIS gives possibility to perform broadband evaluation of ethanol concentration in solution, the employment of resonant cavities allows to achieve very low LOD, laying the groundwork for a new generation of ultrasensitive sensors.